\documentclass[reprint,superscriptaddress,twocolumn,aps,prb,showpacs]{revtex4}
\usepackage{graphicx,float}
\usepackage{amsbsy,amssymb,amsmath,bm,ulem}
\usepackage{color}

\newcommand{\g}{eps}

\begin{document}
\title{Logarithmic relaxation and stress aging in the electron glass}
\author{J. Bergli}
\affiliation{Department of Physics, University of Oslo, P.O.Box 1048
  Blindern, N-0316 Oslo, Norway}
\author{Y. M. Galperin}
\affiliation{Department of Physics, University of Oslo, P.O.Box 1048
  Blindern, N-0316 Oslo, Norway}
\affiliation{A. F. Ioffe Physico-Technical Institute RAS, 194021 St. Petersburg, Russian Federation}
\affiliation{Centre for Advanced Study at the Norwegian Academy of Science and Letters,
0271 Oslo, Norway.}

\begin{abstract}
Slow relaxation and aging of the conductance are experimental features
of a range of materials, which are collectively known as electron
glasses. We report dynamic Monte Carlo simulations of the standard
electron glass lattice model. In a non-equilibrium state, the
electrons will often form a Fermi distribution with an effective
electron temperature higher than the phonon bath temperature. We study
the effective temperature as a function of time in three different
situations: relaxation after a quench from an initial random state,
during driving by an external electric field and during relaxation
after such driving. We observe logarithmic relaxation of the effective
temperature after a quench from a random initial state as well as
after driving the system for some time $t_w$ with a strong electric
field.  For not too strong electric field and not too long $t_w$ we
observe that data for the effective temperature at different waiting
times collapse when plotted as functions of $t/t_w$ -- the so-called
\textit{simple aging}.  During the driving period we study how the
effective temperature is established, separating the contributions
from the sites involved in jumps from those that were not involved. It
is found that the heating mainly affects the sites involved in jumps,
but at strong driving, also the remaining sites are heated.
\end{abstract}
\maketitle

\section{Introduction}

At low temperatures, disordered systems with localized electrons
(e.~g., 
located
 on dopants of compensated doped semiconductors or 
formed by Anderson localization
in disordered conductors) conduct by phonon-assisted
hopping. The theory of this process goes back to Mott\cite{mott} who
invented the concept of variable range hopping (VRH). In particular, he
derived the Mott law for the temperature dependence of the conductance,
\begin{equation}
\sigma \propto \exp [-(\tilde{T}_0/T)^{1/(d+1)}],
\end{equation}
where 
$T$ is temperature, $\tilde{T}_0$ is some characteristic temperature while
$d$ is the dimensionality of the conduction problem. If Coulomb
interactions are important one describes the system as an electron or
Coulomb glass due to the slow dynamics at low
temperatures.\cite{davies}

As is well known (see Ref.~\onlinecite{ES} and references therein),
the single-particle density of states (DOS) develops a soft gap at the
Fermi level, the so called Coulomb
gap.\cite{Pollak70,Pollak71,Srinivasan} While this understanding of
the density of states is generally accepted, the situation is less
clear when it comes to describing dynamics in the interacting case.
Using the Coulomb gap DOS in the VRH arguments\cite{mott} in the same
way as the non-interacting DOS
yields the Efros-Shklovskii (ES) law for
conductance,\cite{ES75,ES76}
\begin{equation}
\sigma  \propto \exp [-(T_0/T)^{1/2}] . 
\end{equation}
This has been observed experimentally in many different types of materials,
like doped semiconductors,\cite{Zabrodskii93,Zhang93,Itoh96,Watanabe98,Massey00}
granular metals \cite{adkins} or two-dimensional systems.\cite{Shlimak,Butko00}

In recent years, there has been increasing interest in the
non-equilibrium dynamics of hopping systems. In particular, the
glass-like behavior at low temperatures has been studied both
experimentally \cite{zvi, ovadyahuPollak, zviStressAging, dragana, grenet, armitage, borini} and
theoretically.\cite{markus,ariel, somoza} In this work, we are
interested in two features 
observed in
the experiments.  Firstly, it was observed
that the conductivity  relaxes logarithmically as a function of time
after an initial quench or perturbation.\cite{ovadyahuPollak} Secondly,
if the system initially in equilibrium is perturbed by some change in
external conditions (e.~g., 
temperature or electric field) for a time
$t_w$ called the waiting time, the relaxation back towards equilibrium
of some quantity like the conductance $G(t,t_w)$ will depend both on
$t_w$ and the time $t$ since the end of the perturbation. It is found
in certain cases that the relaxation is in fact described by a
function $G(t/t_w)$ of the ratio $t/t_w$. This behaviour is 
called \textit{simple aging}.\cite{zviStressAging}

While simple aging is observed in a range of different systems, we are
here particularly concerned with experiments on disordered InO
films.\cite{zvi} One question which has been raised is whether the
observed glassy behavior is an intrinsic feature of the electron
system, or a result of some extrinsic mechanism like ionic
rearrangement.\cite{zviIntrinsic}
In this work, we address the \textit{intrinsic} mechanism by performing
dynamical Monte Carlo simulations
of the standard lattice model of the electron glass. It is known
\cite{somoza} that during a quench from an initial random state 
an effective electron temperature is quickly established, and that
this temperature slowly relaxes to the bath temperature. We show that
the electron temperature relaxes logarithmically over almost three
decades in time and that the system demonstrates simple aging behavior
in a stress aging protocol similar to what is seen in the
experiments.\cite{zviStressAging}  While this does not
constitute a proof of the intrinsic origin of the glassy behavior in
the experiments on InO films, it shows that the model can display the
observed behavior. To the best of our knowledge this was previously
only demonstrated in a mean field approach,\cite{ariel} the accuracy of
which is not well understood.

\section{Model}

We use the standard tight--binding Coulomb glass Hamiltonian,\cite{ES}
\begin{equation}\label{hamil}
H=\sum_i\epsilon_i n_i +\sum_{i<j} {\frac{(n_i-K)(n_j-K)}{r_{ij}}}\;, 
\end{equation}
$K$ being the compensation ratio.  
We take $e^2/d$ as our
unit of energy where $d$ is the lattice constant which we take as our
unit of distance.
The number of electrons is chosen
 to be
half the number of sites so that $K=1/2$.  $\epsilon_i$ are random
site energies chosen uniformly in the interval $[-U,U]$. In the
simulations presented here we used $U=1$, which we know gives a
well-developed Coulomb gap and the ES
law for the 
conductance.\cite{tsigankov, martin, twoElectrons} The sites are
arranged in two dimensions on a $L\times L$ lattice where in all cases
we used $L=1000$, which is sufficiently large to give a good
estimate of the effective temperature in a single state without any
averaging over a set of states.  We implement
cyclic boundary conditions in both directions.

To simulate the time evolution we used the dynamic Monte Carlo method
introduced in Ref.~\onlinecite{tsigankov}. The basic idea of such
simulations is to start the system in some particluar
configuration. The configuration can change by one electron jumping
from site $i$ to site $j$ (in principle, we should also take into
account transitions involving two or more electrons jumping at the
same time, but at the temperatures we consider here this should be a
minor effect, see Ref.~\onlinecite{twoElectrons} and references therein for a
discussion). The energy mismatch between the two states is
supplied by the emission or absorption of a phonon,
therefore
the process is called phonon-assisted tunneling.

The rate of a phonon-assisted transition from site $i$ to site $j$ 
is usually specified as (see, e.g., 
Ref. \onlinecite{ES}),

\begin{equation}\label{gamma}
\Gamma_{ij} \propto |\gamma_q|^2|N(\Delta E_{ij})|e^{-2 r_{ij}/a} \, .
\end{equation}
Here $\Delta E_{ij}$ is the phonon energy, $r_{ij} \equiv
|\mathbf{r}_i -\mathbf{r}_j|$, $a$ is the localization length of the
electronic state which we choose to 1, while $\gamma_q$ is a coupling
constant, which in general depends on the wave vector $q$ of the
involved phonon. Since $q \propto |\Delta E_{ij}|$ the pre-factor
$\gamma_q$ leads to a power-law dependence of the transition rate on
$|\Delta E_{ij}|$ where the power is model dependent.  Since the
power-law pre-exponential dependence does not change the results in a
qualitative way we assume $\gamma_q$ to be $q$-independent.  $N(E) =
(e^{E/T}-1)^{-1}$ is the equilibrium phonon density. We set $k_B=1$ so
that temperatures and energies are measured in the same units.

In the given initial state, one should, in principle, calculate all such
rates, then select one at random weighted by the rates. The selected
transition is then performed, all Coulomb energies recalculated, and
the process repeated from the new state. Note that the rates depend on
the states through the energy changes $\Delta E_{ij}$, which include
the contributions from the Coulomb energy. All the rates therefore have
to be recalculated at each step. In practice, one restricts the
possible jumps by only considering those where the distance $r_{ij}$
is less than some maximal length (in our simulations we only
considered jumps of less than 10 lattice units) since longer jumps
become so improbable that they are never selected anyway. The
procedure of selecting the next jump is also optimized in other ways,
see Ref.~\onlinecite{tsigankov} and, in particular, 
Ref.~\onlinecite{tsigankovPRB} for  a description of how to calculate
the proper elapsed time. 
The only difference from Ref.~\onlinecite{tsigankov} is that following
Ref.~\onlinecite{tenelsen}, we use for the transition 
rate $\Gamma_{ij}$ 
 instead of (\ref{gamma}) the approximate formula
\begin{equation} \label{eq:01}
\Gamma_{ij} = {\tau_0}^{-1}e^{-2 r_{ij}/a}\min 
  \left(e^{-\Delta E_{ij}/T},1\right)
\end{equation}
where $\Delta E_{ij}$ is the energy of the phonon and $r_{ij}$ is the
distance between the sites.  $\tau_0$ contains material dependent
factors and energy dependent factors, which we approximate by their
average value; we consider it as constant and its value, of the order
of $10^{-12}$ s, is chosen as our unit of time.  (Note that in
Ref. \onlinecite{tsigankov} a different approximate formula was used,
we do not believe that the difference is of great significance,
although it may change numerical values).

\section{Relaxation and effective temperature}

To get more detailed understanding of how the effective temperature is
established, we can follow the effective temperature as a function of
time after an initial quench or some external perturbation like a
strong electric field is applied. 

\begin{figure}[t]
\begin{center}
\includegraphics[width=\linewidth]{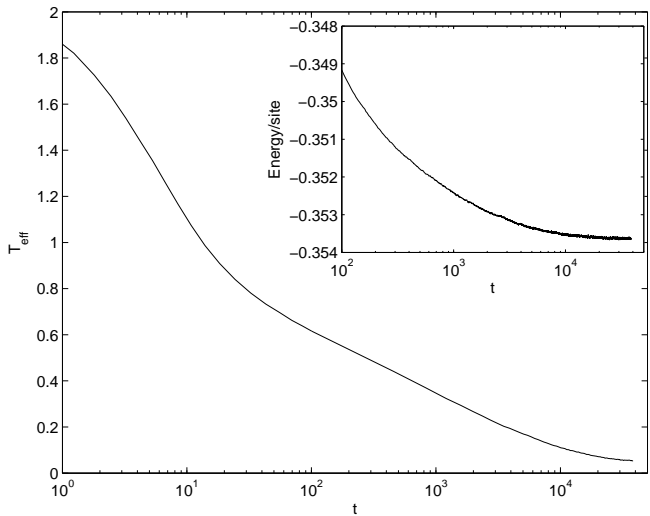}
\caption{\label{fig:TeffofT} Effective temperature as function of time. The inset shows the energy as function of time.  } 
\end{center}
\end{figure}

Let us first relax from an initial random state and measure
$T_{\text{eff}}(t)$. In all our simulations we used  a phonon
temperature $T=0.05$, which we know is well into the ES
regime for VRH.
The graphs show the evolution over $10^8$
jumps. 
Shown in  Fig.~\ref{fig:TeffofT} are the energy (inset) and the
effective temperature as functions of time. 

As we can see, the energy graph has almost stopped to decrease,
indicating that we have almost reached equilibrium. The same is seen
by the effective temperature, where $T_{\text{eff}}= 0.054$ in the
final state. We see that the effective temperature, after some initial
short time, logarithmically decreases in time for about two and a half
orders of magnitude.  The energy does not show this behavior (as
discussed in Ref.~\onlinecite{martinRelax}, it is well fitted by a
stretched exponential function). The relaxation of the effective
temperature was studied previously\cite{somoza} using a similar
numerical method, but having the sites at random instead of on a
lattice. A lower temperature was also used, and together this slows
down the simulations so that only much smaller samples, up to 2000
sites, could be studied. Instead of out linear fit to effective
temperature as a function of $\ln t$ they obtain a linear fit to
effective temperature as a function of $1/\ln t$. The reason for this
difference is not clear, but it may result from our larger samples and
longer times.
\begin{figure}[t]
\begin{center}
\includegraphics[width=\linewidth]{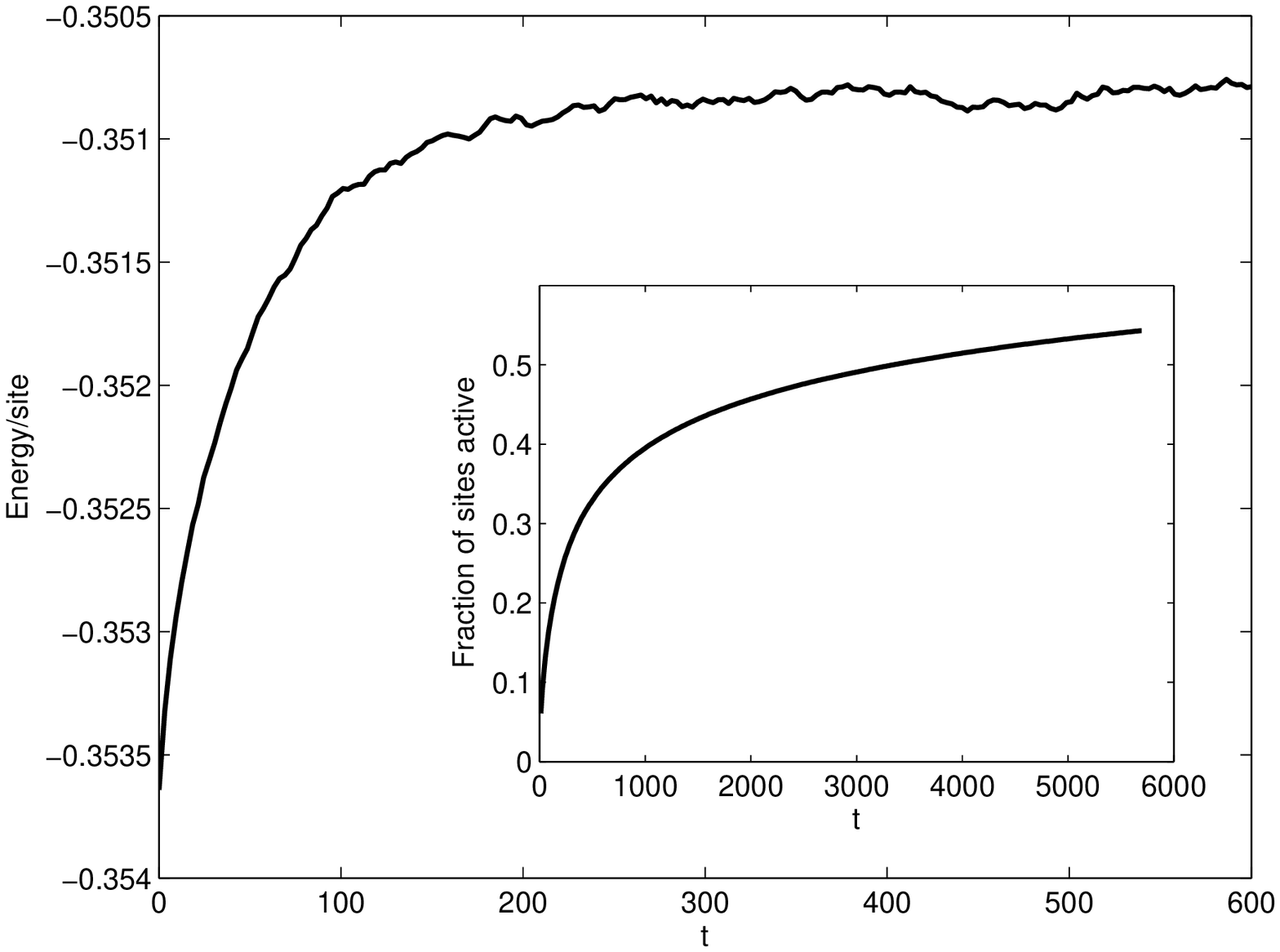} \\
\includegraphics[width=\linewidth]{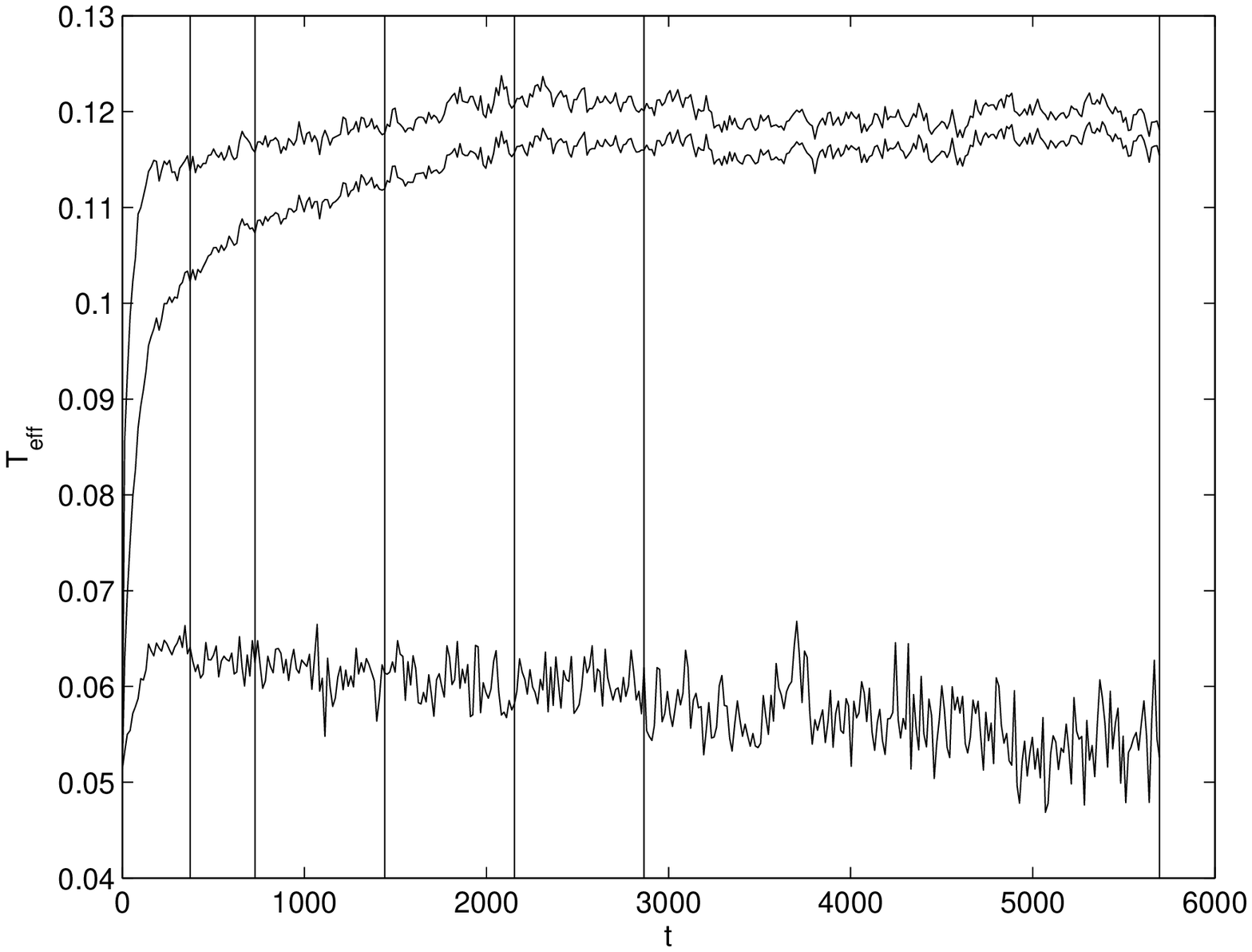}
\end{center}
\caption{\label{fig:TeffEM} Top: Energy as function of time. Inset:
  The fraction of sites involved in a jump as function of time.  Bottom:
  Effective temperature as function of time. The curves are from top
  to bottom the temperature of the sites which were active, the
  temperature of all sites  and the temperature of the sites which were
  not active. The vertical lines indicate the waiting times in the
  stress aging protocol (see Fig. \ref{fig:stressAging}) .
}
\end{figure}

Another type of experiment is to relax the system at low temperature
and then apply a strong electric field to pump energy into the
system. To
equilibrate the system, we simulated the system with a localization
length of 100, which facilitates long jumps and faster
equilibration. When the energy did not decrease any more, we switched
to the normal localization length of 1 and observed that the energy
remained constant with fluctuations and $T_{\text{eff}}=0.05\pm0.001$ after
each $10^6$ jumps for a total of $7\cdot10^6$ jumps.
We then applied an electric field $E=0.1$
(in units of $e/d^2$).
We know that Ohmic
conduction takes place when $E\lesssim T/10$, so this should be well
into the non-Ohmic regime and we expect the electron temperature to
increase above the phonon temperature. 
 Shown in the top panel of  Fig. \ref{fig:TeffEM} is the energy per site 
as a function of time.  The effective temperature as a function of time 
is shown in the lower panel (middle curve).  Noting the difference in the timescales
we conclude
 that the energy stabilizes at a new value much faster than the effective
temperature. 

We have also plotted the effective temperature taking into account
only those sites which were involved in a jump, Fig. \ref{fig:TeffEM}
(bottom, upper curve), and those which did not jump,
Fig. \ref{fig:TeffEM} (bottom, lower curve). As we can see, the sites
which are not involved in jumps are still at a temperature close to
the phonon temperature.  This is not a trivial statement, since the
energies of the sites which are not involved in jumps also change due
to the modified Coulomb interactions with the sites that jumped. Note
that even at the latest time shown, new sites are still being
involved, see Fig. \ref{fig:TeffEM} (top, inset), even if the energy and
effective temperature are more or less stable.

\begin{figure}[h]
\begin{center}
\includegraphics[width=0.32\linewidth]{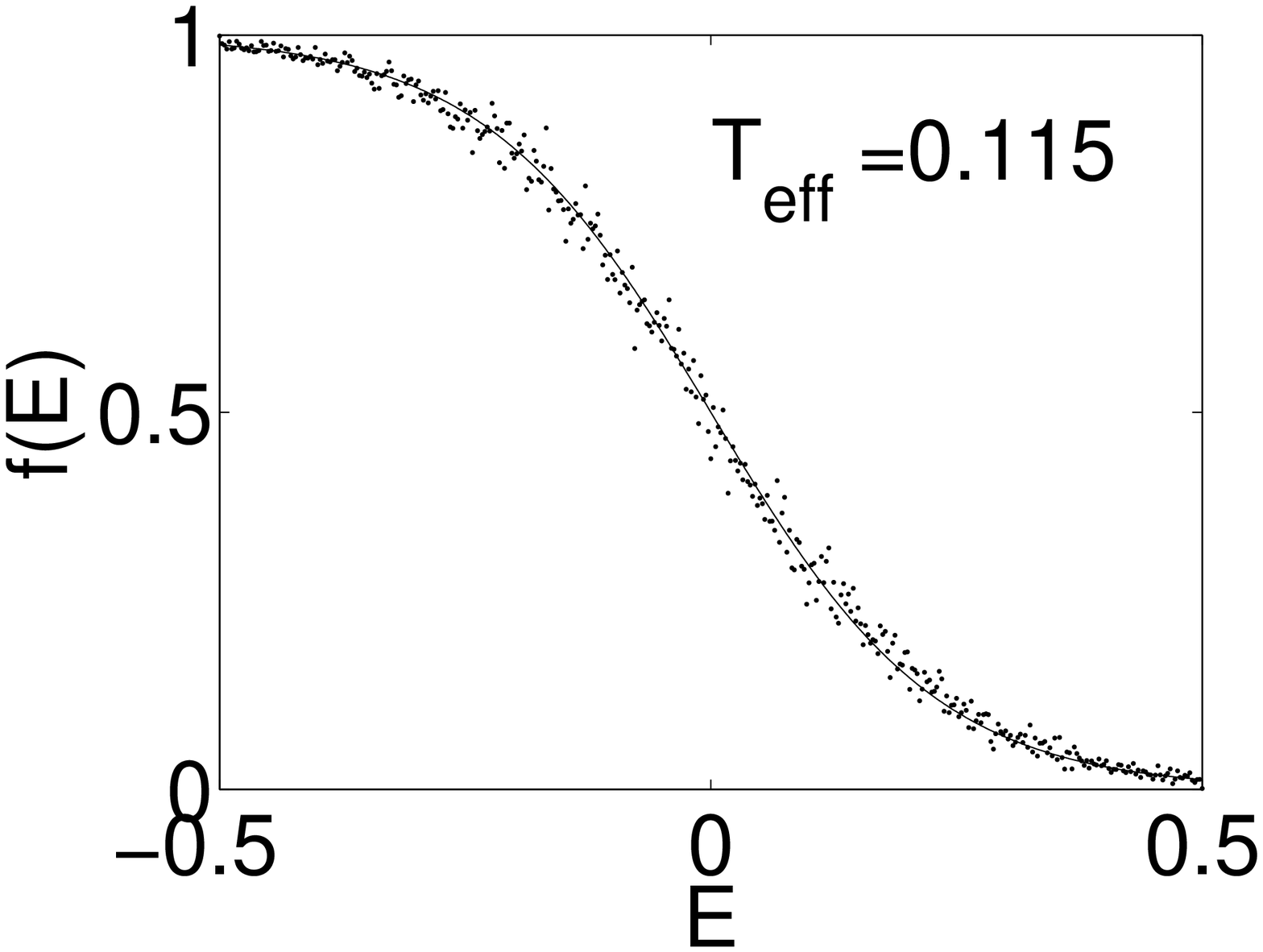} \hfill
\includegraphics[width=0.32\linewidth]{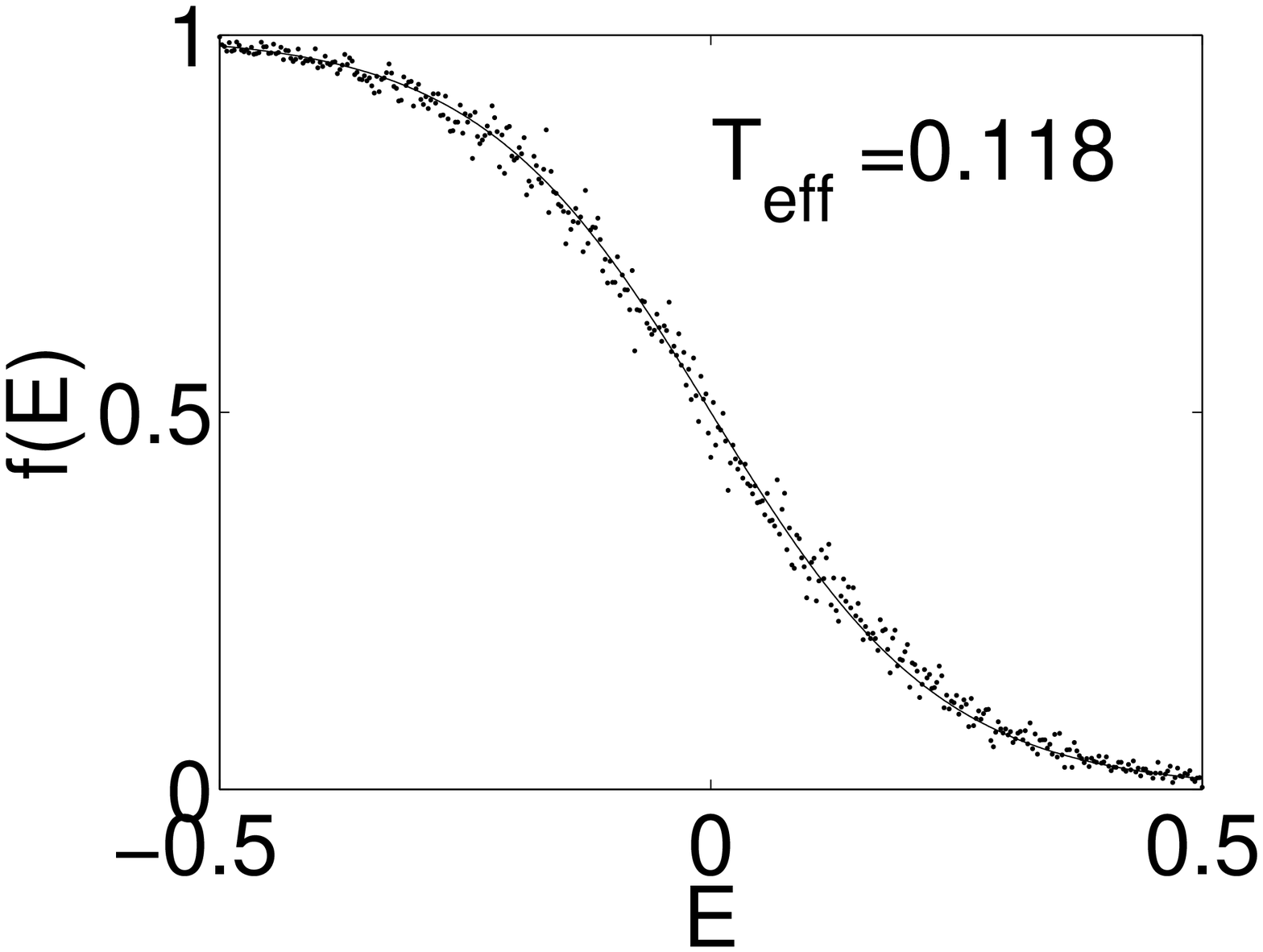} \hfill
\includegraphics[width=0.32\linewidth]{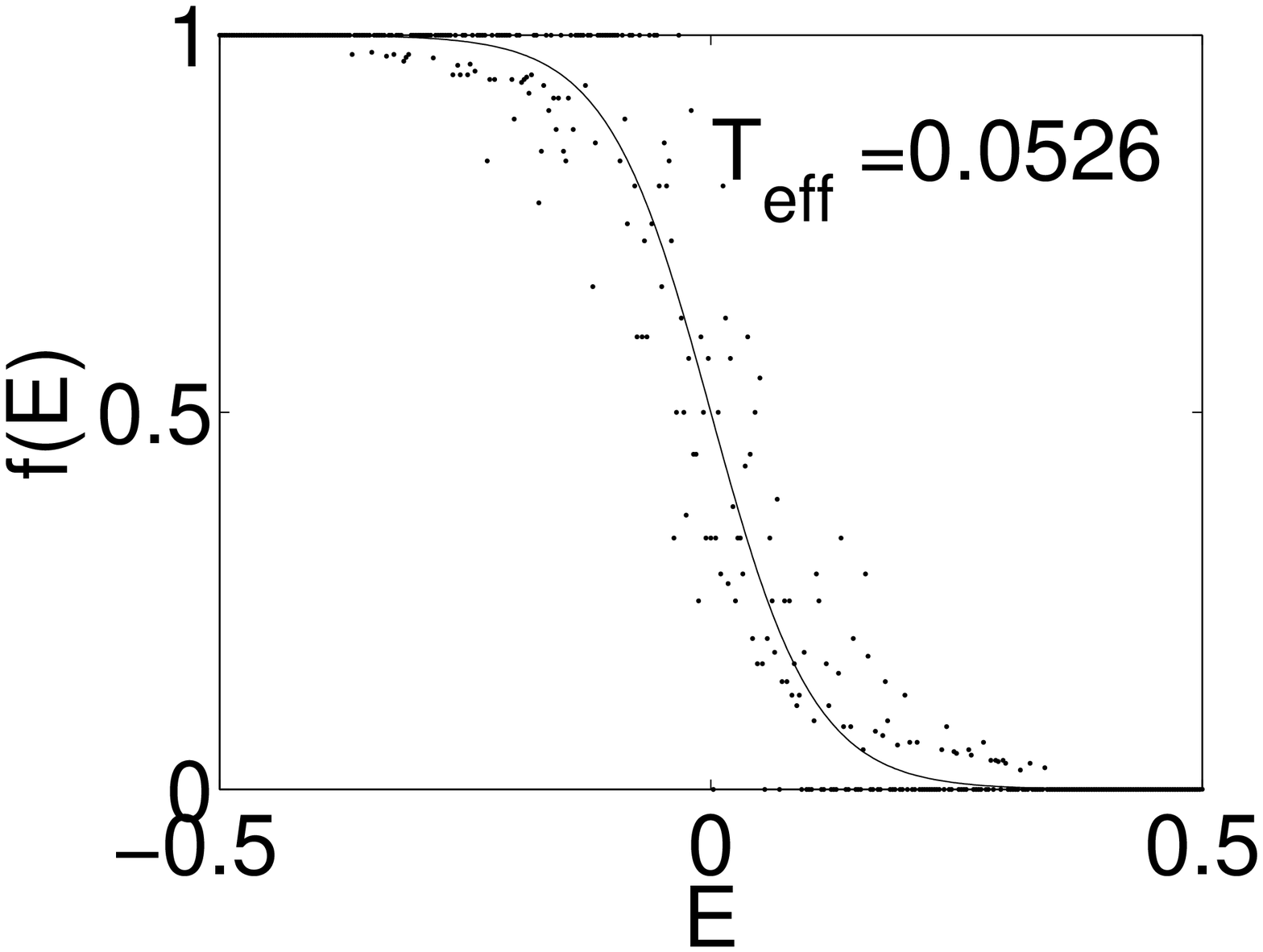}
\end{center}
\caption{\label{fig:Fermi}Fermi functions. Left: All sites, Center:
  Sites which took part in a transition, Right:  Sites which did not
  take part in a transition. The points are the data and the curve
  the fitted function. } 
\end{figure}

The fitted Fermi functions are shown in Fig.~\ref{fig:Fermi}.
\begin{figure}[t]
\begin{center}
\includegraphics[width=0.99\linewidth]{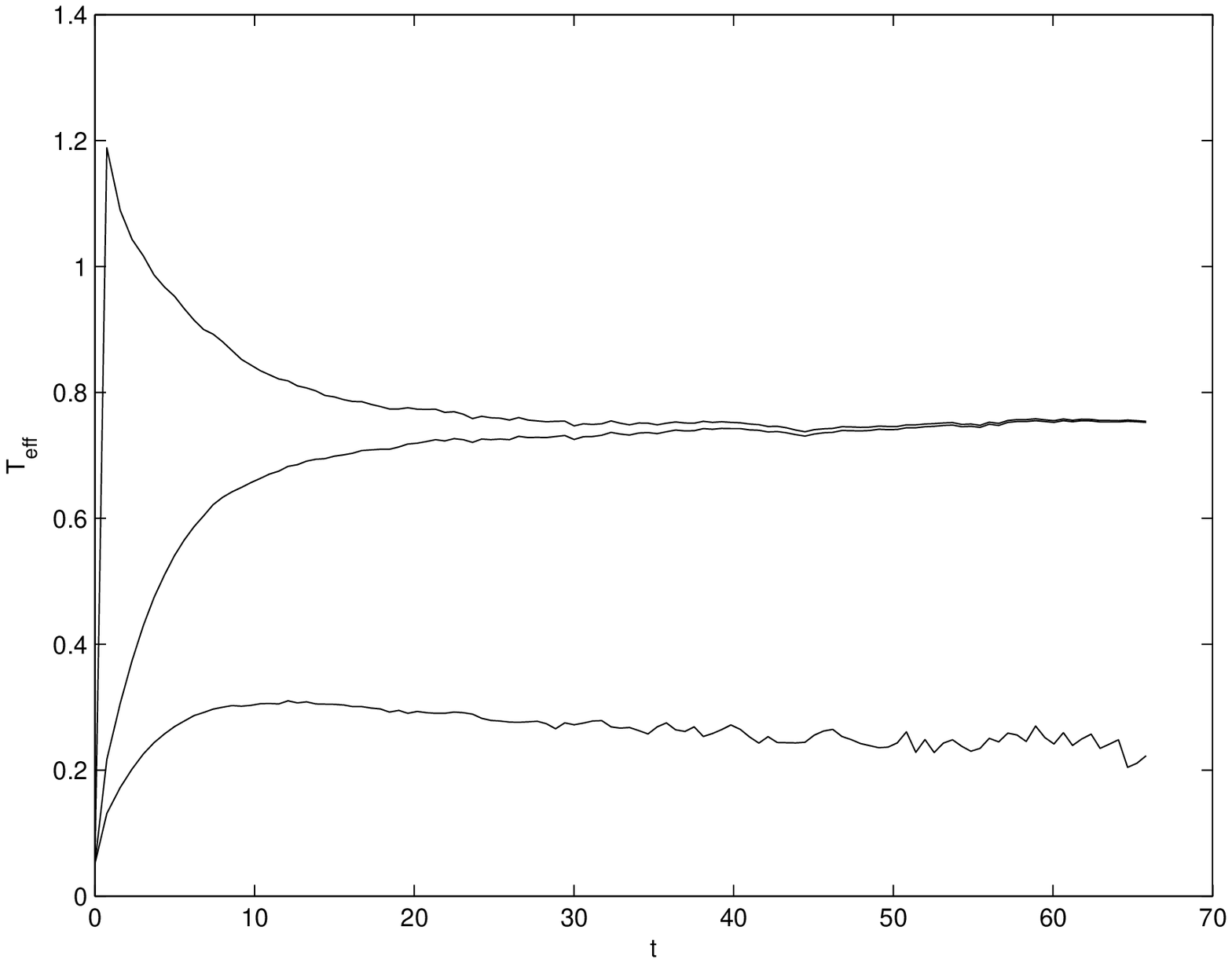} \\ 
\includegraphics[width=0.32\linewidth]{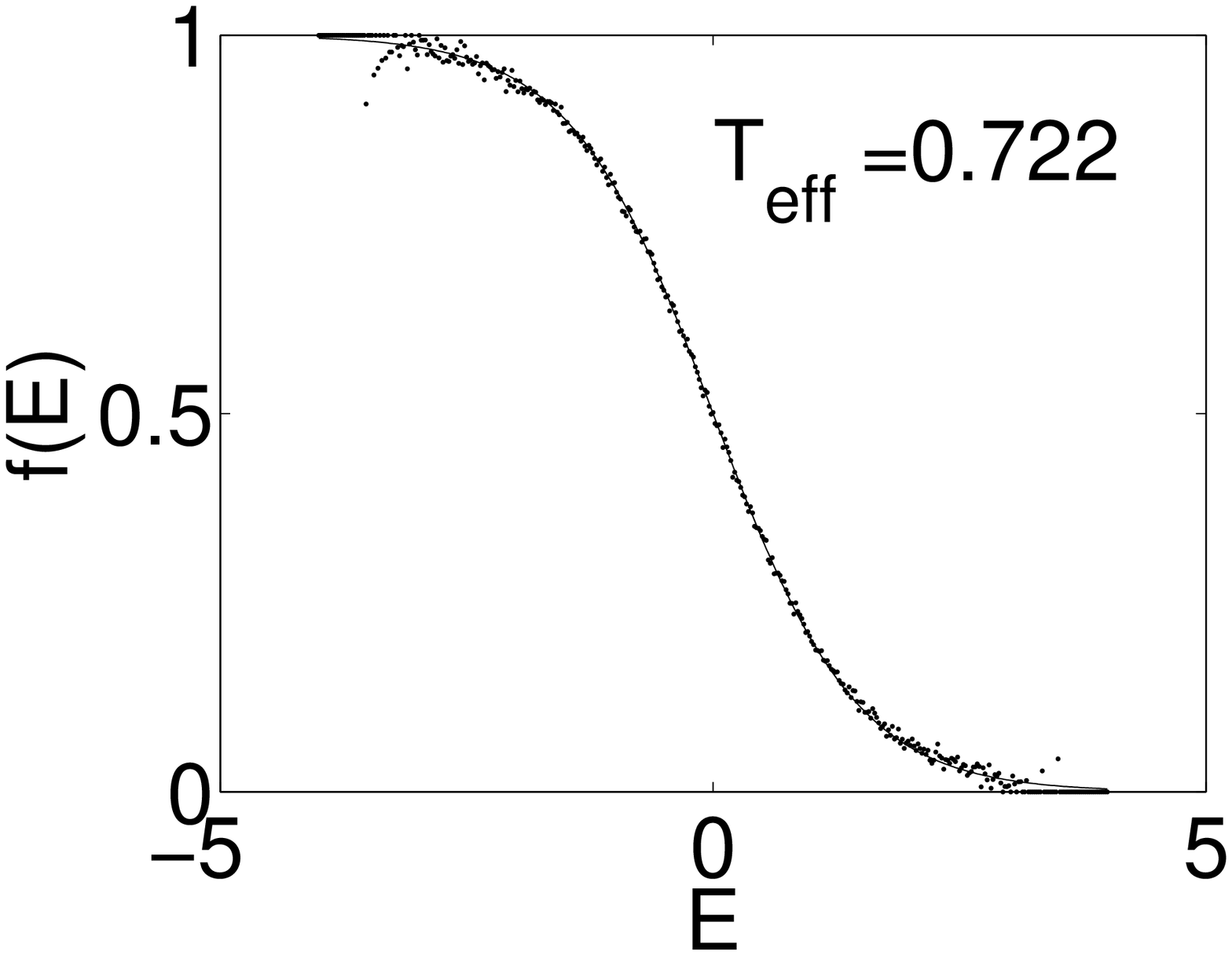} \hfill
\includegraphics[width=0.32\linewidth]{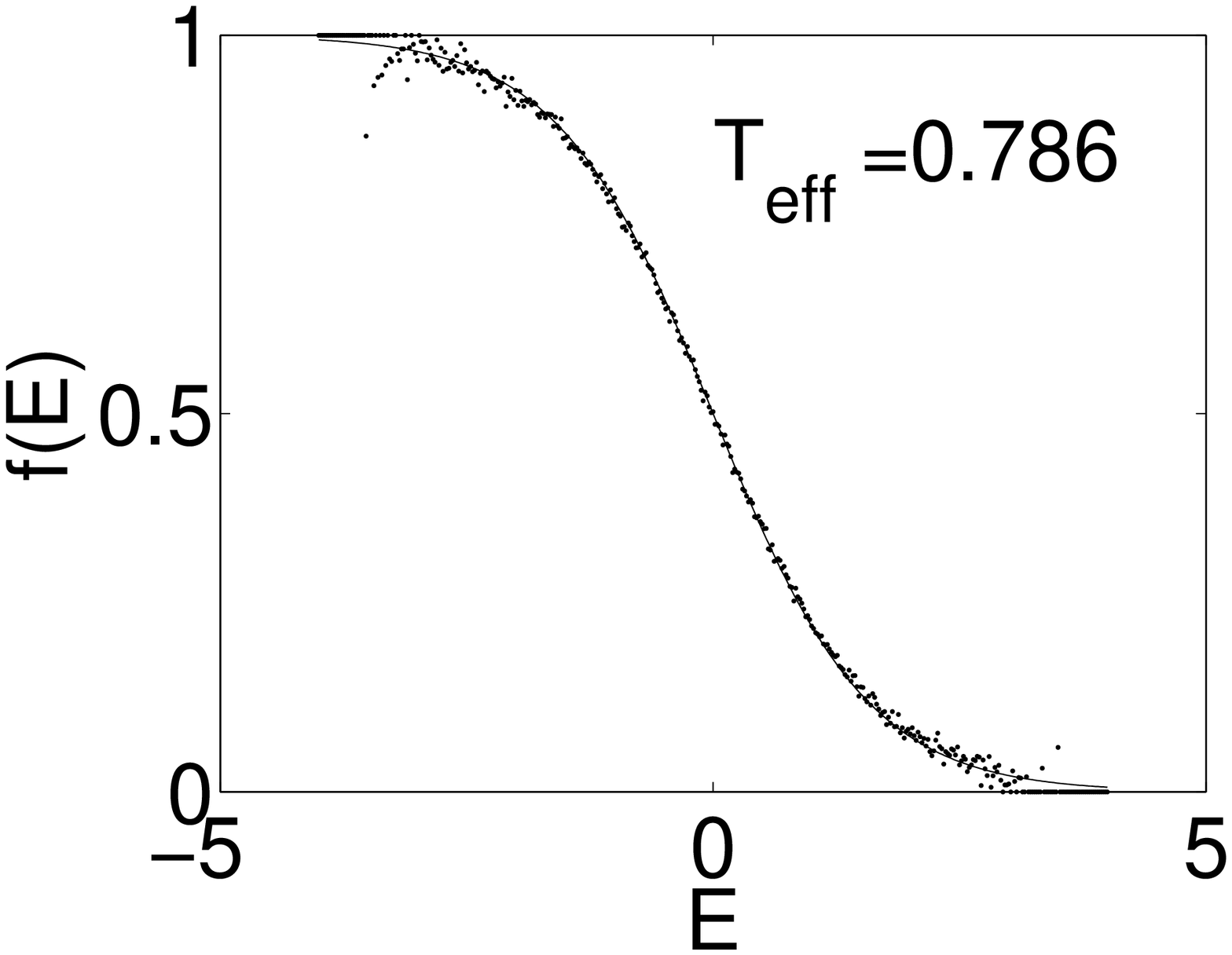} \hfill
\includegraphics[width=0.32\linewidth]{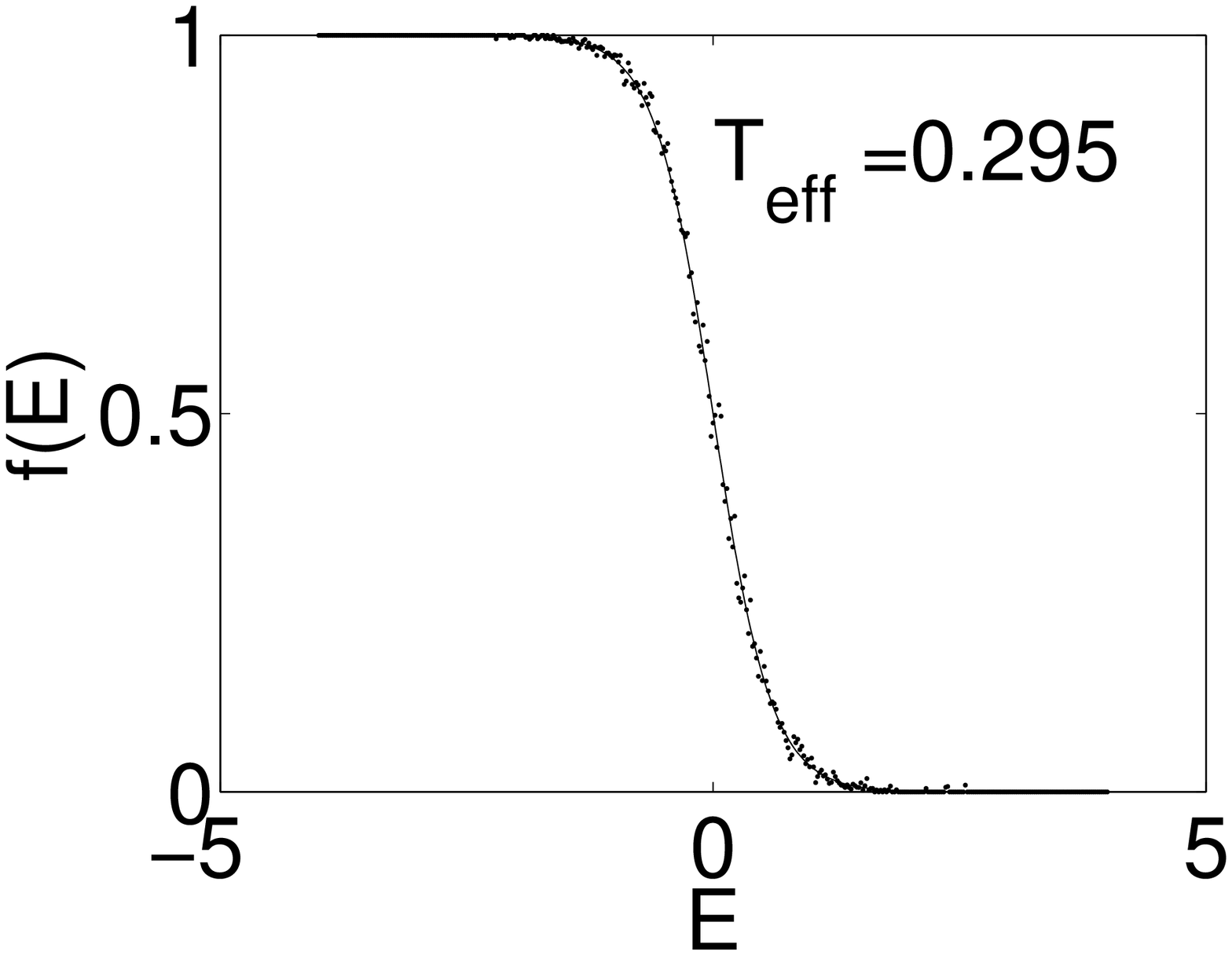}
\end{center}
\caption{\label{fig:TeffE1}
When driving with electric field
$E=1$. Top: $T_{\text{eff}}$ as function of time. The curves are from top to
bottom the temperature of the sites which were active, the
temperature of all sites and the temperature of the sites which
were not active.  Bottom: Fitted Fermi functions at time
$t=19$. Left: All sites, Center: Sites which took part in a
transition, Right: Sites which did not take part in a transition. The points are the data and the curve
  the fitted function. }
\end{figure}
The fits are quite good, but the data for the sites which never were
involved in jumps are more noisy. Note that the number of sites which
jumped is still not much more than half the total number of sites, so
the noise is not because there are few sites, but rather (as could be
expected) that the sites which did not jump are those which are far
from the Fermi level. These are either filled or empty and do not 
contribute much to the effective temperature. The noise indicates that
the energy shifts are not correlated with the original energy of the
site, so that there is no systematic change in the occupation
probability giving a change in the effective temperature. 

If we drive with a stronger field $E=1$ (Fig.~\ref{fig:TeffE1}), we
observe two new features: First, the temperature of the sites which
took part in a jump changes non-monotonously, first increasing rapidly
and then decreasing towards a stationary value. Note however that at
very short times, before the peak is reached, the distribution is not
close to a Fermi distribution and the temperature is not a meaningful
concept. On the decreasing part of the curve the Fermi function was
already established. Second, we now observe significant heating also
of the sites, which did not take part in a jump. Because of the Coulomb
interaction it is not surprising that also these sites are affected by
the transitions on other sites. What is surprising is that the
occupation numbers still follow the Fermi distribution quite
closely. Notice that the data are not noisy like in the case of weaker
driving (Fig.~\ref{fig:Fermi}), so the shifts in the energy levels are
systematically adjusting to the Fermi distribution.

\section{Stress Aging}

Motivated by the logarithmic relaxation of effective temperature in
Fig.~\ref{fig:TeffofT} and using the heating curve of Fig~\ref{fig:TeffEM} 
we will try to reproduce the
experimental results\cite{zviStressAging} using the stress
aging protocol. This means applying a non-Ohmic field for a certain
time $t_w$ and then turning this field off (keeping only an Ohmic
measuring field, which supposedly does not appreciably perturb the
system). The heating process of Fig.~\ref{fig:TeffEM} is exactly such a
non-Ohmic driving, and we need only start simulations with zero fields
at different points along this curve (in the experiments they kept an
Ohmic measuring field, but since we are monitoring the effective
temperature not the current this is not needed. We also ran
simulations in Ohmic fields and confirmed that this did not
appreciably affect the effective temperature as was expected). We have
choosen six $t_w$, which correspond to the points marked on
Fig.~\ref{fig:TeffEM} with vertical lines.

\begin{figure}[t]
\begin{center}
\includegraphics[width=0.99\linewidth]{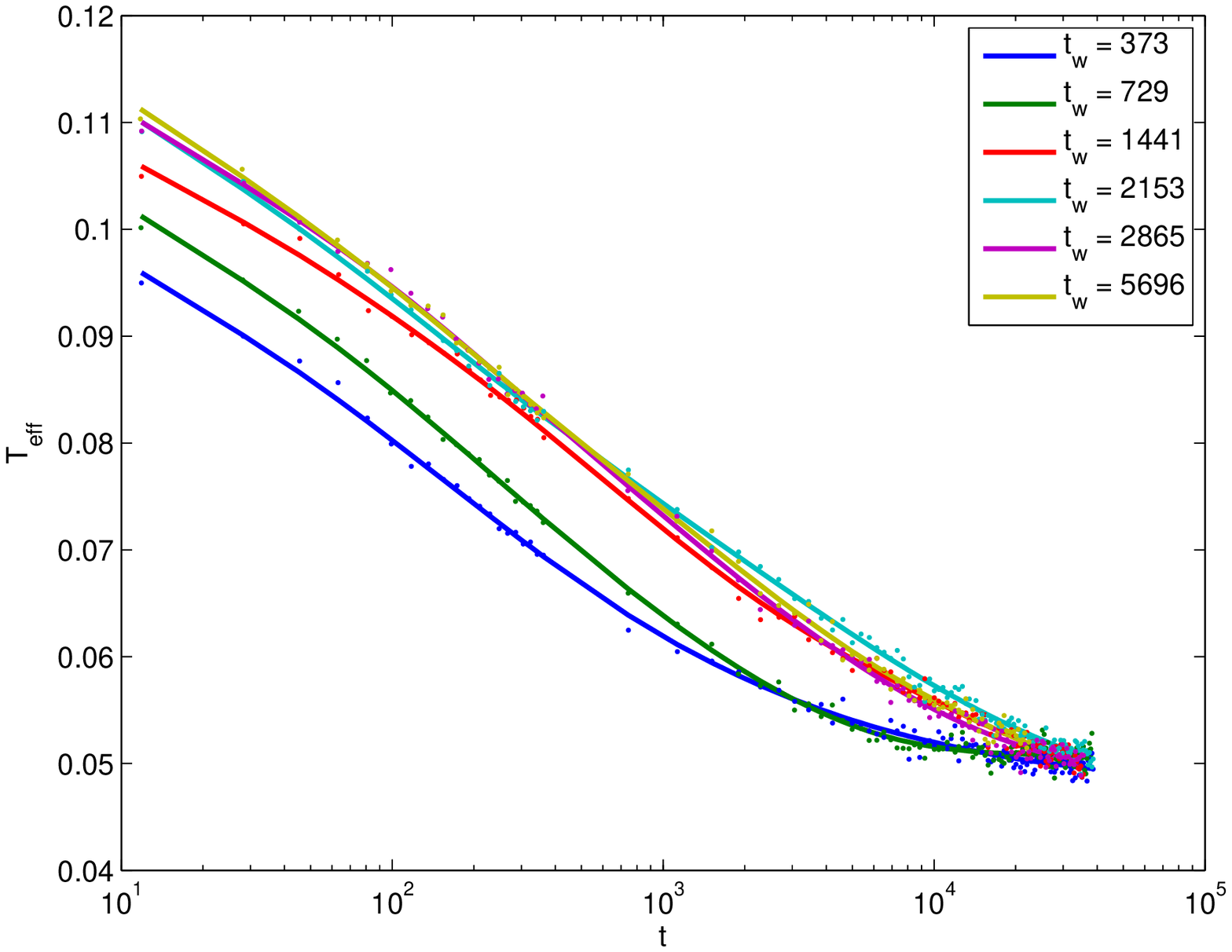}
\includegraphics[width=0.99\linewidth]{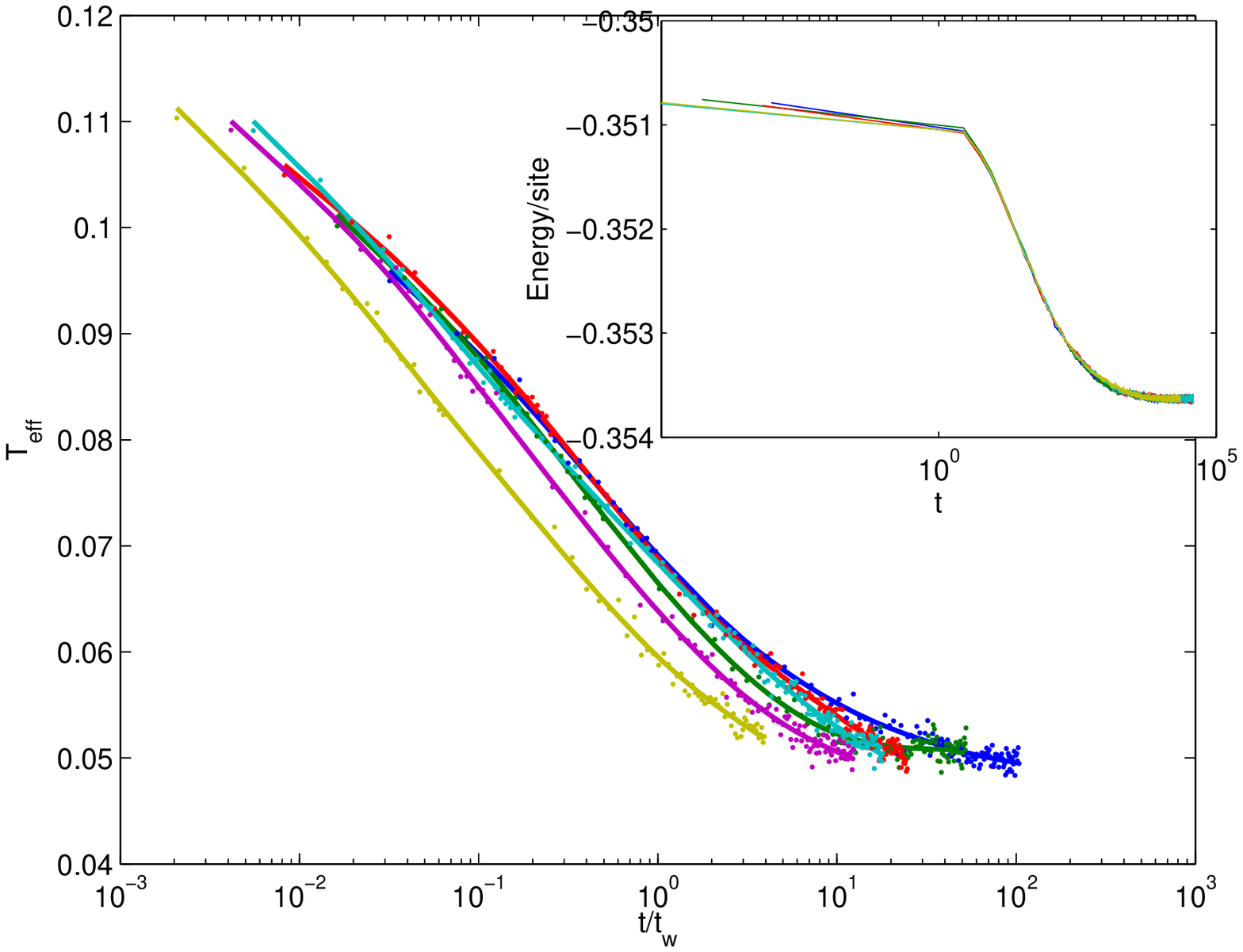}
\end{center}
\caption{\label{fig:stressAging} (Color online) After driving with $E=0.1$. Top:
$T_{\text{eff}}$ as function of time.  Bottom: $T_{\text{eff}}$ as function of
$t/t_w$. Inset: Time dependences of the energy for different $t_w$.} 
\end{figure}

Figure \ref{fig:stressAging} (top) shows the
effective temperature as function of time after the end of the driving period. 
Note that the time dependence of the energy is very similar in all
cases as shown in Fig.~\ref{fig:stressAging} (bottom, inset). The
effective temperature as a function of $t/t_w$ is shown in
Fig.~\ref{fig:stressAging} (bottom).

From Fig.~\ref{fig:stressAging} we conclude that there is a
logarithmic relaxation of the effective temperature after a driving by
a non-Ohmic field just as in the case of relaxation from a random
initial state (Fig.~\ref{fig:TeffofT}). Furthermore, we see that the
curves for different $t_w$ collapse when time is scaled with $t_w$
when $t_w$ is smaller than some critical value $t_w^{(c)} \approx 2500$.
The curve for $t_w=2865$ seems to lie a little
to the left of the collapse curve, and for $t_w=5696$ this tendency is
clear. The collpase of the curves for short $t_w$ is similar to what
is observed in the experiments both on indium oxide
films\cite{zviStressAging} and porous silicon.\cite{borini} In the
case of porous silicon, also the departure from the simple aging at
longer waiting times was observed, while sufficiently long times were
never reached in the case of indium oxide. If we
compare to Fig.~\ref{fig:TeffEM} (bottom) we see that the critical
value $t_{w}^{(c)}$ corresponds to the time where the effective
temperature stabilizes. Comparing to Fig.~\ref{fig:TeffEM} (top) we
see that this is a time much longer than the one, which is needed for
the energy to stabilize.

To check that this behavior is not particular to one specific sample
we repeated the procedure (relaxing to equilibrium using large
localization length, then the stress aging protocol) on a different
sample which gave similar results.

\begin{figure}
\begin{center}
\includegraphics[width=0.99\linewidth]{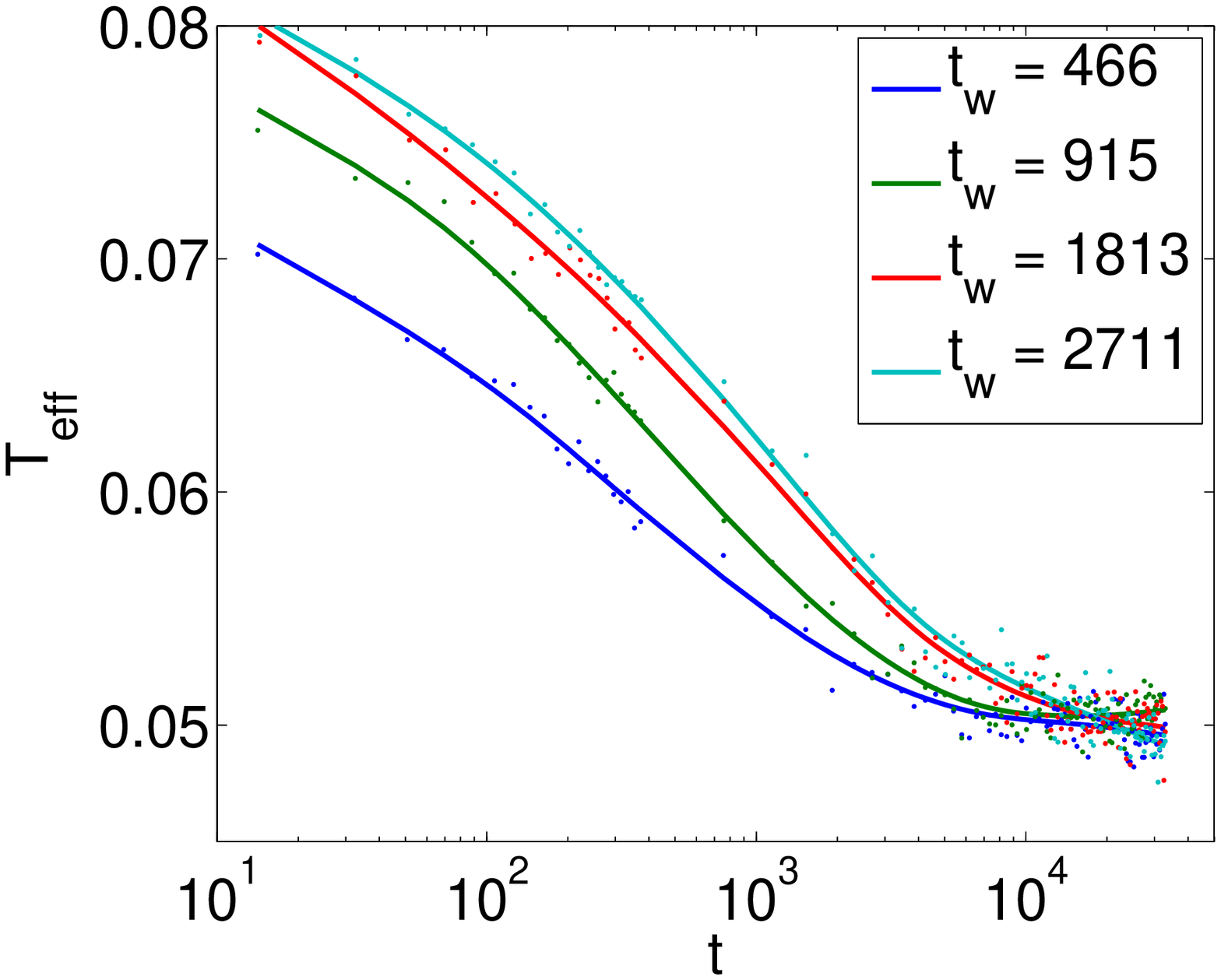} \hfill
\includegraphics[width=0.99\linewidth]{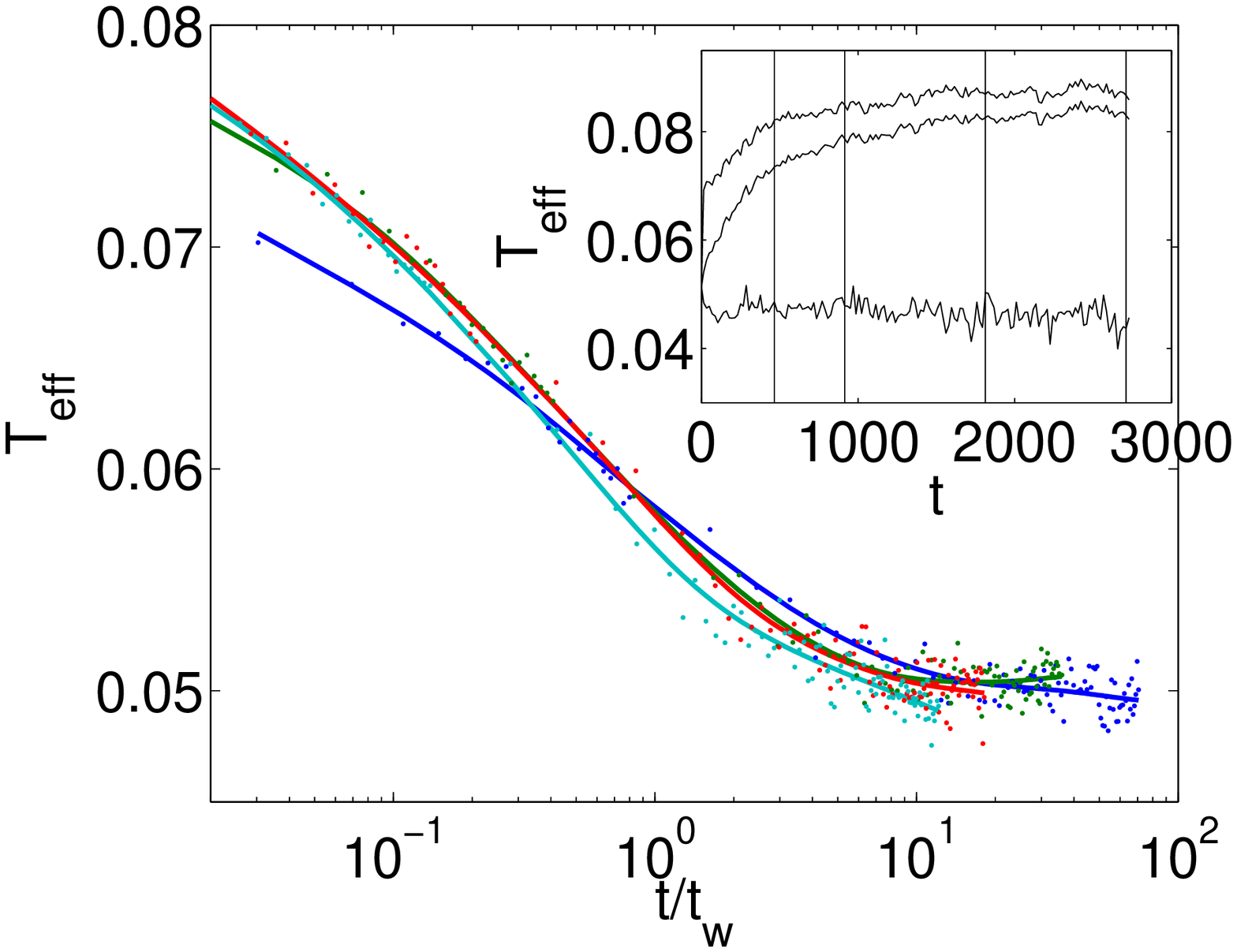}
\caption{\label{fig:stressAgingE0.05} (Color online) After driving with $E=0.05$. Top: $T_{\text{eff}}$ as function of time.  Bottom:
$T_{\text{eff}}$ as function of $t/t_w$. 
Inset: The heating curve.}
\end{center}
\end{figure}

\begin{figure}
\begin{center}
\includegraphics[width=0.99\linewidth]{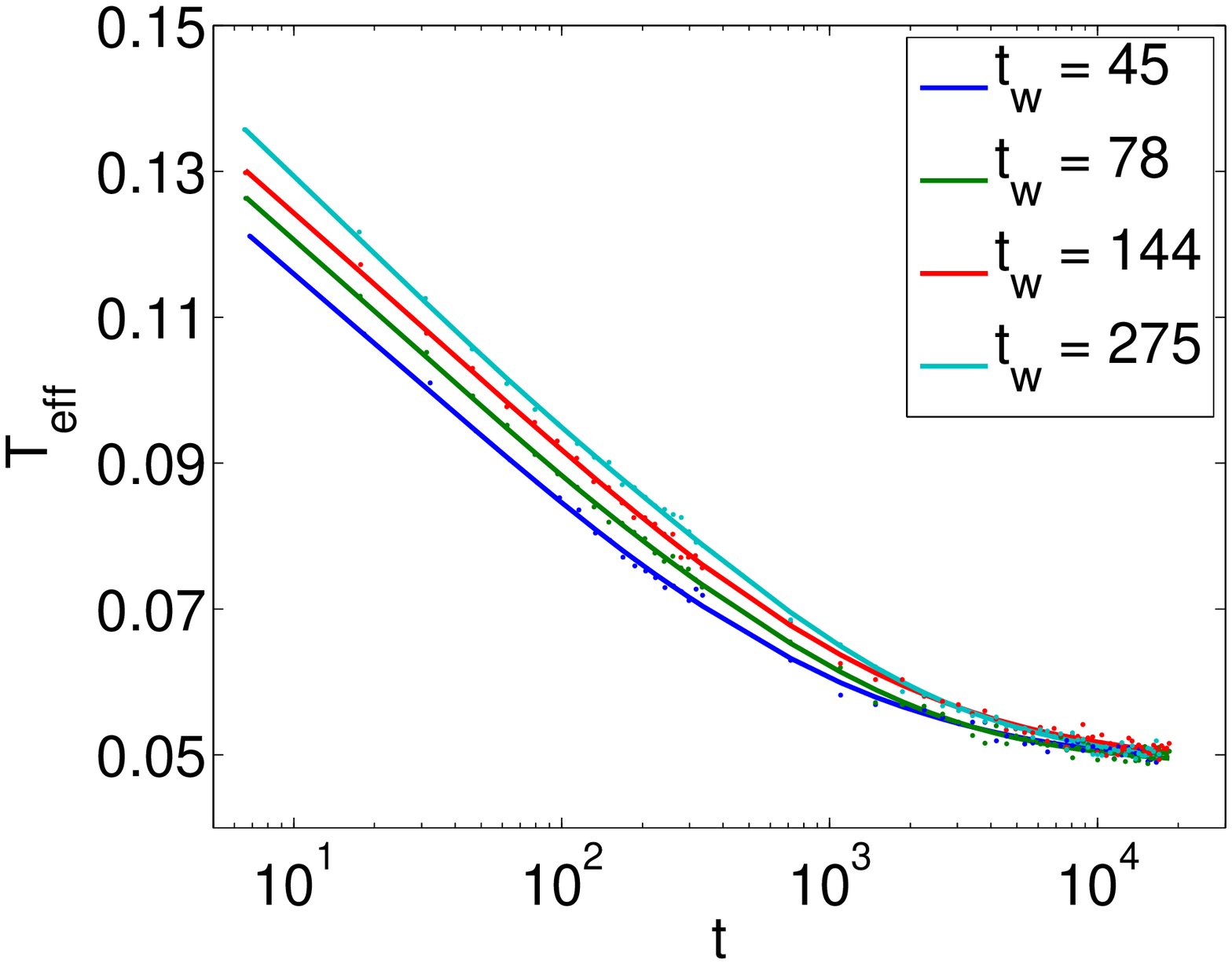} \hfill
\includegraphics[width=0.99\linewidth]{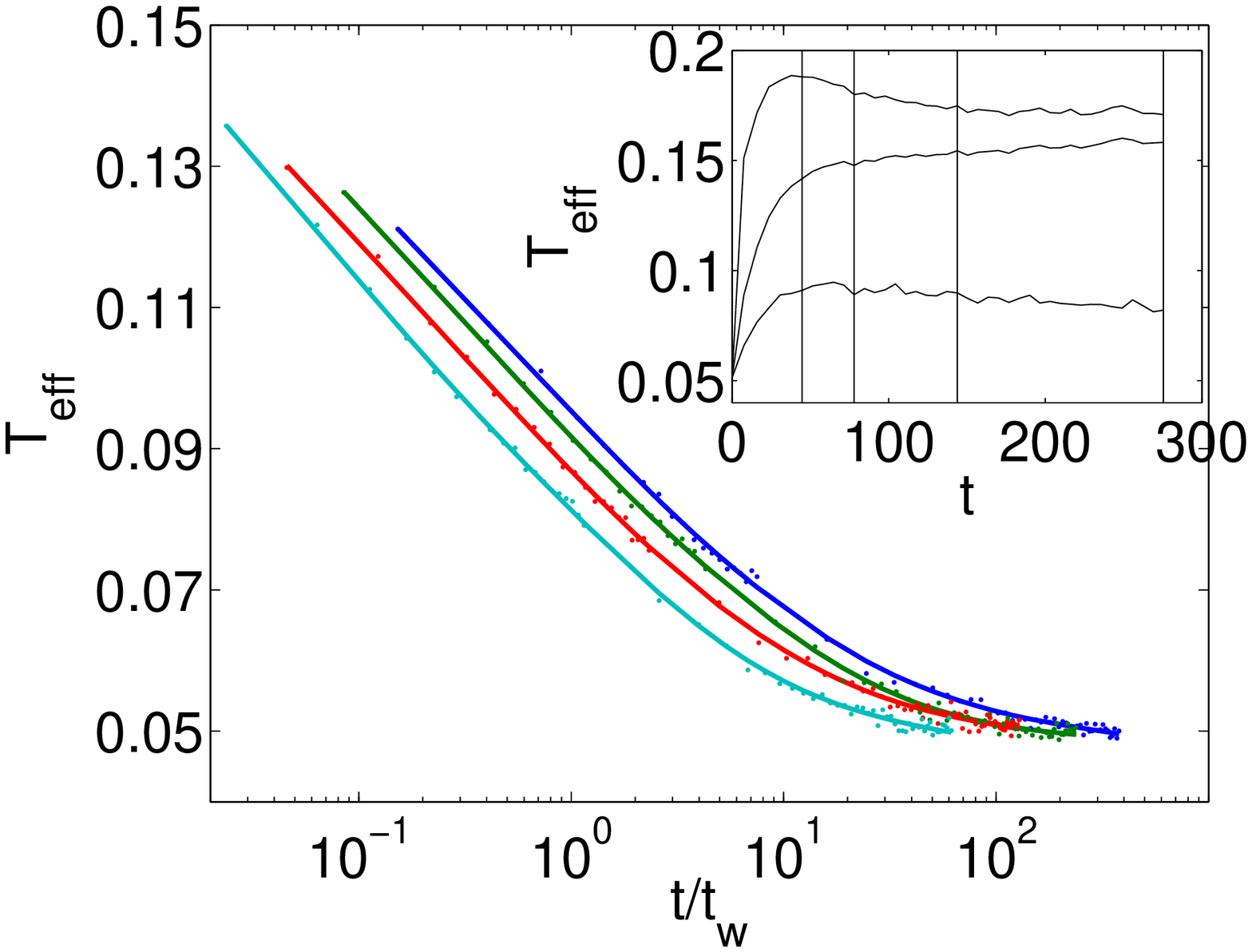}
\end{center}
\caption{\label{fig:stressAgingE0.2} (Color online) After driving with $E=0.2$. Top: $T_{\text{eff}}$ as function of time.  Bottom:
$T_{\text{eff}}$ as function of $t/t_w$.  Inset: The heating curve.}
\end{figure}

For the first sample we also repeated the stress aging protocol at
different driving fields, $E=0.05$, 0.2, 0.5 and 1. Note that to be in
the Ohmic regime we should have $E\lesssim T/10$, so all the fields are
well outside of this. For $E=0.05$ the
heating curve is shown in Fig.~\ref{fig:stressAgingE0.05} (inset)
and the effective temperature as function of time after the end of
 the driving period is shown in  Fig.~\ref{fig:stressAgingE0.05}.  
As we see, the general behavior is the same as when driving with
$E=0.1$. 
 For $E=0.2$ the
heating curve is shown in Fig.~\ref{fig:stressAgingE0.2} (inset)
 and the effective temperature as function of time after the end of
 the driving period is shown in  Fig.~\ref{fig:stressAgingE0.2}.  
We see that the curves do not collapse satisfactorily, even for $t_w$
shorter than the time at which $T_{\text{eff}}$ stabilizes. This 
behavior is also observed  in the experiments,\cite{zviStressAging} 
it is characteristic for stronger fields.

Overall, what we observe is qualitatively very close to what is seen
in experiments, but it should be noted that in experiments it is
always the conductance that is measured, whereas we have studied the
effective temperature. If we believe that the nonlinearity of the
conductance is mainly due to heating of the electrons, then the
nonlinear conductance can be found as the linear conductance,
$\sigma_0$, at the effective temperature. It is known that this is
approximately true.\cite{caravaca} Based on this assumption and
expecting that $\sigma_0(T_{\text{eff}})$ is an analytical function
one would expect that $\Delta \sigma \equiv \sigma - \sigma_0(T)
\propto T_{\text{eff}} -T$ at $T_{\text{eff}} -T \ll T$.  This would
be sufficient to show that the behavior of the conductivity would be
the same as what we see for the effective temperature.

However, in our case $T_{\text{eff}}$ is significantly greater than $T$, 
and the above consideration seemingly does not work. Therefore we
made an attempt to study conductivity directly. Instead of
turning off the field after $t_w$ we switched to $E=0.005$ which
should be more or less the highest field which is still in the Ohmic
range.  The effective temperature shows behavior which is similar to
the one seen at $E=0$, which is what we expect since this field is so
weak as to hardly affect the effective temperature.

To 
simulate
the relaxation of conductivity is not so easy. Firstly,
because to find the conductance, the system has to be followed over
some time and transported charge measured. Usually it is noisy and one
needs to average over a considerable time. With a large system like
the one we have used, it is
better, but still difficult to follow changes in the current. Secondly,
and more importantly, when the strong, non-Ohmic field is applied,
the system is polarized by charges shifting locally. These are charges
which are not on the conducting path, and  therefore they do not
contribute to the DC current. But when the field is switched to a
small, non-Ohmic value, these charges will flow backwards. Since we
find the current by counting transferred charge in the direction of
the field for each jump this will lead to a negative current for some
time until this polarization has relaxed. 
One can ask whether this
would also be seen in real experiments which only measures current in
an external circuit. It seems that even if the localized charges are
not on the conducting path, they would be capacitively coupled to the
external circuit, and thereby detectable in real experiments. No such
effect has been reported. 

To see the relaxation of the conductivity, we took the accumulated
transferred charge and subtracted the same in the absence of measuring
field, which should contain only the relaxation of the
polarization. The result was smoothed and the derivative calculated to
find the current. The resulting curve showed some relaxation of the
current, but the noise was too large and further work is needed in
order to draw any clear conclusions.

\section{Discussion}

The phenomenology observed in our simulations agrees with what is seen
in the experiments \cite{zviStressAging,borini} in virtually all essential
aspects: We observe logarithmic relaxation of the effective
temperature after a quench from a random initial state. We also
observe logarithmic relaxation after driving the system for some time
$t_w$ with a strong electric field. When the driving field is not too
strong and $t_w$ not too long we observe simple aging in the sense
that the curves for different waiting times collapse on a common curve
when plotted as functions of $t/t_w$. When $t_w$ exceeds some critical
value, $t_{w}^{(c)}$, the scaled curves do not follow the common curve.
When the driving field is large the curves do not collapse even for
short times. The only difference is that while in the experiments
conductance was measured, we have studied the effective temperature.
We have to rely on the assumption that the conductance is given by
the linear conductance at the effective temperature to relate our
results to the experiments. 

\begin{figure}
\begin{center}
\includegraphics[width=\linewidth]{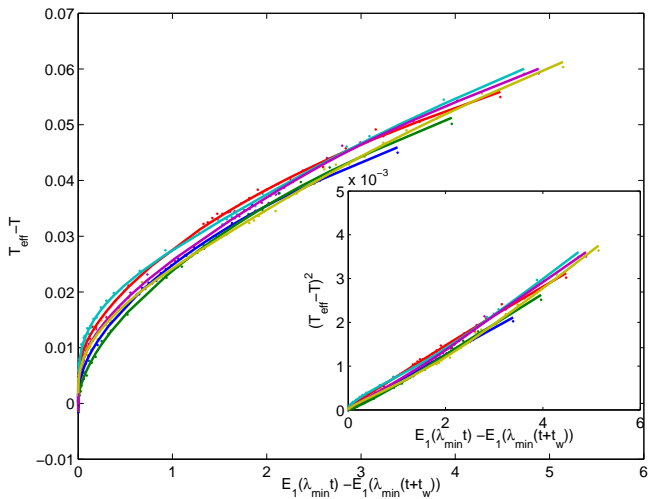}
\caption{ (Color online) $\delta T$ as function of $E_1(\lambda_{\min}t) -
E_1[\lambda_{\min}(t+t_w)]$. Inset: $\delta T^2$ as function of
$E_1(\lambda_{\min}t) - E_1[\lambda_{\min}(t+t_w)]$
after driving with $E=0.1$.
The data are the same
as in Fig. \ref{fig:stressAging} and the color labels are the same.\label{fig:meanField}}
\end{center}
\end{figure}

The aging protocol discussed here is the subject of a recent mean
field analysis.\cite{ariel,borini} It predicts the
relaxation of the  average occupation number after the field is turned
off. It is then argued that the excess current  
should follow the same law,
\[
 \delta\sigma  
  \sim \int_{\lambda_{\min}}^{\lambda_{\max}}\frac{d\lambda}{\lambda}
   \left(1-e^{-\lambda t_w}\right)e^{-\lambda t_w}\, ,
\]
where $\lambda_{\min}$ and  $\lambda_{\max}$ are the slowest and fastest
modes. If we assume that $\lambda_{\max}t\gg1$ we can set
$\lambda_{\max}=\infty$ and we get
\begin{equation}\label{eq:ariel}
 \delta\sigma  
  \sim E_1(\lambda_{\min}t) -  E_1[\lambda_{\min}(t+t_w)]
\end{equation}
where $E_1(x) = \int_x^\infty t^{-1}e^{-t}\, dt$ is the exponential integral
function. Its series expansion is $E_1(x) = -\ln x -\gamma + x
+\ldots$ so that for $t,t_w\ll/\lambda_{\min}$ we get $\delta\sigma
\sim \ln(1+t_w/t)$. When $t\ll t_w$ it reduces to what we have used
before. The idea is that the failure to collapse the curves for long
$t_w$ is because $\lambda_{\min}t_w\gtrsim1$ and we have to use the
full exponential integral instead of only the leading logarithmic
term. Assuming the excess effective temperature follows the same law
as the conductance we should plot $\delta T = T_{\text{eff}}-T$ as function
of $E_1(\lambda_{\min}t) -  E_1[\lambda_{\min}(t+t_w)]$ and obtain a
collapse of the curves to a straight line
(Fig. \ref{fig:meanField}). This was indeed observed in the
experiments on porous silicon.\cite{borini}
$\lambda_{\min}$ is now a fitting parameter, which we fit by hand to
find the best collapse when $\lambda_{\min}=2.5\cdot10^{-4}$. This
agrees at least approximately with the idea that
$1/\lambda_{\min}=4000$ should be 
not far from
the $t_{w}^{(c)}=2500$ that
we found above. 
The collapse is not to a straight line, but rather close to a square
root. We can see this clearer if we plot (Fig. \ref{fig:meanField}, inset)
$\delta T^2$ as function of $E_1(\lambda_{\min}t) -
E_1[\lambda_{\min}(t+t_w)]$. 
All the curves fall close to the straight line indicating that the
scaling relation predicted by the mean field theory is respected by
our data. However, we can ask why $\delta T^2$ rather than $\delta T$
fall on a straight line. If we believe that our numerical model is
applicable to the experiments, it seems to indicate that
$\delta\sigma\propto\delta T^2$ instead of $\delta\sigma\propto\delta
T$ as we obtained above from the assumption that the conductance is the
linear conductance at the effective temperature  and $\delta T/T \ll 1$. 

\section{Conclusions}

We have demonstrated logarithmic relaxation of the effective
temperature after a quench from a random initial state. The same is
observed after driving by some non-Ohmic electric field. In the latter
case we also observe simple aging when the driving field is not too
strong and the waiting time not too long. At longer waiting times or
after driving with stronger electric fields we observe departure from
the simple aging qualitatively similar to what is seen in
experiments.

The Monte Carlo approach allows us to access several properties, which
are not available either in the experiments or the mean field theory.
When applying a non-Ohmic field both the average energy and the
effective temperature increase and saturate at a level above the
equilibrium one. We find that the saturation of energy is much faster
than the saturation of effective temperature.

 We also see that the
heating mainly affects those sites which were involved in jumps. At
moderate driving fields the
energies of the remaining sites are shifted by the changing Coulomb
interactions, but there is no sytematic shifts, and the best fitting
Fermi distribution is still at or close to the bath temperature. 
At stronger fields there is also some heating of the sites which were
not involved in jumps, and the distribution still follows a Fermi
function.


\begin{thebibliography}{99}

\bibitem{mott}
N. F. Mott, J. Non-Cryst. Solids {\bf 1}, 1 (1968).

\bibitem{davies}
J. H. Davies, P. A. Lee, and T. M. Rice, Phys. Rev. Lett. {\bf 49}, 758
(1982).


\bibitem{ES}
B. I. Shklovskii and A. L. Efros, Electronic properties of doped
semiconductors (Springer, Berlin, 1984).

\bibitem{Pollak70} M. Pollak, Discuss. Faraday Soc. \textbf{50}, 13 (1970).
\bibitem{Pollak71}  M. Pollak, Proc.
R. Soc. London, Ser. A \textbf{325}, 383 (1971).
\bibitem{Srinivasan} G. Srinivasan, Phys. Rev. B \textbf{4}, 2581 (1971).

\bibitem{ES75} A. L. Efros and B. I. Shklovskii, J. Phys. C \textbf{8}, L49 (1975). 

\bibitem{ES76}  A. L. Efros, J. Phys. C \textbf{9}, 2021 (1976).

\bibitem{Zabrodskii93}A. G. Zabrodskii and A. G. Andreev,  JETP Lett.,
\textbf{58}, 756 (1993).

\bibitem{Zhang93}J. Zhang, \textit{et al.},
 Phys. Rev. B, \textbf{48}, 2312 (1993).

\bibitem{Itoh96} K. M. Itoh, \textit{et al.},
 Phys. Rev. Lett., \textbf{77}, 4058 (1996).

\bibitem{Watanabe98}M. Watanabe, \textit{et al.},
Phys. Rev. B, \textbf{58}, 9851 (1998).

\bibitem{Massey00}J. G. Massey and M. Lee,  Phys. Rev. B, \textbf{62},
R13270 (2000).

\bibitem{adkins}C. J. Adkins, J. Phys.: Cond. Matt. {\bf 1},
1253 (1989).

\bibitem{Shlimak}I. Shlimak, \textit{et al.},
Phys. Rev. B, \textbf{61}, 7253 (2000).

\bibitem{Butko00} V. Y. Butko, J. F. DiTusa and P. W. Adams,
Phys. Rev. Lett. \textbf{84}, 1543  (2000).

\bibitem{zvi} A. Vaknin, Z. Ovadyahu, and M. Pollak,
 Phys. Rev. Lett. {\bf 84}, 3402 (2000); Phys. Rev. B, {\bf 65},
 134208 (2002).

\bibitem{ovadyahuPollak}
Z. Ovadyahu and M. Pollak, Phys. Rev. B {\bf 68}, 184204 (2003).

\bibitem{zviStressAging}
V. Orlyanchik and Z. Ovadyahu, Phys. Rev. Lett. {\bf 92}, 066801
(2004). 

\bibitem{dragana}
S. Bogdanovich and D. Popovi\'c, Phys. Rev. Lett. {\bf 88}, 236401 (2002);
J. Jaroszy\'nski and D. Popovi\'c, Phys. Rev. Lett. {\bf 96}, 037403 (2006).

\bibitem{grenet}
T. Grenet , J. Delahaye, M. Sabra, and F. Gay,
Eur. Phys. J. B {\bf 56}, 183 (2007).

\bibitem{armitage} 
V. K. Thorsm\o lle and N. P. Armitage,
Phys. Rev. Lett. {\bf 105}, 086601 (2010).

\bibitem{borini}
Ariel Amir, Stefano Borini, Yuval Oreg, and Yoseph Imry,
Phys. Rev. Lett. {\bf 107}, 186407 (2011).

\bibitem{markus}
M. M\"uller and L. B. Ioffe, Phys. Rev. Lett. {\bf 93}, 256403 (2004).
Eran Lebanon and Markus M\"uller, Phys. Rev. B {\bf 72}, 174202 (2005).


\bibitem{ariel}
Ariel Amir, Yuval Oreg, and Yoseph Imry,
Phys. Rev. Lett. {\bf 103}, 126403 (2009). 

\bibitem{somoza}
A. M. Somoza, M. Ortu\~no, M. Caravaca, and M. Pollak
Phys. Rev. Lett. {\bf 101}, 056601 (2008).

\bibitem{zviIntrinsic}
Z. Ovadyahu,  Phys. Rev. B {\bf  78}, 195120 (2008).


\bibitem{tsigankov}
D. N. Tsigankov and A.~L.~Efros, Phys. Rev. Lett. {\bf 88}, 176602 (2002).

\bibitem{martin}
A. Glatz, V. M. Vinokur, J. Bergli, M. Kirkengen, and Y.~M.~Galperin
J. Stat. Mech.  P06006 (2008). 

\bibitem{twoElectrons}
J. Bergli, A. M. Somoza, and M. Ortu\~no
Phys. Rev. B {\bf 84}, 174201 (2011). 

\bibitem{tsigankovPRB}
D. N. Tsigankov, E. Pazy, B. D. Laikhtman, and A.~L.~Efros, Phys. Rev. B, {\bf 68} 184205 (2003).

\bibitem{tenelsen}
K. Tenelsen and M. Schreiber, Phys. Rev. B {\bf 52}, 13287 (1995);
A. D\'iaz-S\'anchez et. al., Phys. Rev. B {\bf 59}, 910 (1999).

\bibitem{martinRelax}
M. Kirkengen and J. Bergli
Phys. Rev. B {\bf 79}, 075205 (2009). 

\bibitem{caravaca}
M. Caravaca, A. M. Somoza, and M. Ortu\~no, 
Phys. Rev. B {\bf 82}, 134204 (2010).

\end{thebibliography}
\end{document}